\documentclass[12pt]{article}
\usepackage[letterpaper, left=0.75in, right=0.75in, top=0.75in, bottom=0.75in]{geometry}
\usepackage{graphicx} % Required for inserting images
\usepackage{caption}
\usepackage{amsmath}
\usepackage{subcaption}
\usepackage{booktabs}
\usepackage[sort&compress, numbers]{natbib}
\usepackage{authblk}

\title{Reducing entanglement with a Hamiltonian derived Clifford transformation}

\author[1]{James Brown}
\author[2]{Erika Lloyd}
\author[2]{Alexandre Fleury}
\author[1]{Tarini S Hardikar}
\author[1]{Kenny Heitritter}

\affil[1]{qBraid Co., Chicago, IL, USA}
\affil[2]{SandboxAQ, Palo Alto, CA, USA}
\date{\today}

\begin{document}

\maketitle
\begin{abstract}
    Recently (Physica Scripta, 100(10):105401, 2025), an algorithm was introduced that deterministically generates a Clifford transformation from the Qubit Coupled Cluster (QCC) algorithm which we call Q-Cliff (QCC+Clifford). There, it was shown that Q-Cliff could be utilized to generate a hardware efficient version of the QCC ansatz. Here, we examine and refine these techniques and show that Q-Cliff can be utilized to generate efficient classical and quantum approximations to the ground states of chemical systems. The algorithm generates an efficient variational method that generally has accuracy between MP2 and CISD with $\mathcal{O}(N^6)$. Furthermore, we show through DMRG calculations that the entanglement between qubits is reduced significantly and therefore the accuracy for a given bond dimension can be vastly improved (up to an order of magnitude). Finally, we refine the previously reported algorithm to generate low-depth and CNOT efficient circuits that can be optimized with a comparable number of energy evaluations to state-of-the-art VQE algorithms. All these results show that this Hamiltonian derived Clifford transformation should be a tool used for many classical and quantum algorithms.
\end{abstract}

\section{Introduction}
One of the most promising applications of quantum computers is quantum chemistry\cite{WEIDMAN2024102105}, as one can leverage the inherent quantum mechanical nature of quantum computers to simulate molecular systems that are intractable for classical computers. While classical computers struggle with the exponential scaling of quantum many-body problems, quantum computers naturally handle this exponential scaling, which makes them particularly well-suited for tackling complex chemical phenomena like bond breaking, catalysis, and electronic correlation effects. Quantum algorithms could eventually enable the design of new materials, drugs, and catalysts by providing accurate solutions to problems that would require prohibitively long computation times on even the most powerful supercomputers.

Quantum advantage in chemistry lies in algorithms such as the Variational Quantum Eigensolver\cite{Peruzzo_2014} (VQE) and Quantum Phase Estimation\cite{nielsen2010quantum} (QPE), which potentially can find ground-state energies and explore chemical reaction pathways. However, any advantage depends on the effective preparation of a good approximation to the desired quantum state\cite{nielsen2010quantum}. Poor state preparation can severely compromise the efficiency and accuracy of quantum chemistry algorithms. If the initial quantum state is far from the true ground state of a molecule, the quantum algorithm may require many more iterations to converge, potentially negating any quantum advantage. Moreover, on current noisy intermediate-scale quantum (NISQ) devices, where quantum coherence is limited, starting with a good initial guess becomes even more crucial to obtain meaningful results before noise destroys the quantum information.

Sophisticated state preparation techniques, such as adiabatic preparation, variational state preparation, and chemistry-inspired ans\"{a}tze\cite{Anand_2022, Ryabinkin_2018} that incorporate known molecular orbital structures, are therefore essential for bridging the gap between theoretical quantum advantage and practical quantum chemistry applications. The quality of state preparation often determines whether a quantum chemistry calculation succeeds or fails, making it one of the most critical components in the quantum computing toolkit for molecular simulation.

Recently\cite{brown2025iterative}, a method that can produce short-depth quantum circuits and initial parameters derived from Qubit Coupled Cluster \cite{ryabinkin2018qubitcoupledclustermethodsystematic, Ryabinkin_2018} (QCC) was introduced which we refer to as Q-Cliff. This circuit can be further optimized using hybrid methods or can also be utilized in the future for QPE or other early fault-tolerant algorithms\cite{Katabarwa_2024}. In this manuscript, we examine the technique in more detail and show that the single layer (zero correlation ansatz) provides a cheap approximation for chemical systems that is variational with accuracy (relative to FCI) approximately between MP2 and CCSD. By utilizing DMRG calculations, we also show that entanglement is drastically reduced between the qubits and pushes the entanglement to fewer qubits. We further compare the full ansatz to the recently introduced CEO-ADAPT-VQE algorithm\cite{Ram_a_2025} which has drastically reduced metrics (number of measurements, CNOT gates, circuit depth) compared to other varational quantum algorithms. We show that Q-Cliff can generate accurate energies in VQE using fewer measurements, fewer two-qubit gates and lower depth circuits for many systems. 

\section{Q-Cliff: The QCC derived Clifford Transformation}

Q-Cliff begins with the QCC ansatz. This takes a Hamiltonian described as a linear combination of Pauli words $P_{i}$ which are products of $I$, $X$, $Y$ and $Z$ Pauli operators. This can be written as 
\begin{equation}\label{eq:pauli_ham}
    H = \sum_{i}c_iP_i
\end{equation}
which for the quantum chemistry problems we examine could be obtained by any standard  fermion-to-qubit mappings\cite{JKMN,Setia_2019,Bringewatt2023parallelization} from the fermionic Hamiltonian. Although we focus here on chemistry applications, QCC can be applied to any Hamiltonian of the form of Eq. \ref{eq:pauli_ham}.

Each of these Pauli words is then grouped using their ``flip'' indices, which are the qubits that have either a $X$ or $Y$ operator. For example, a Pauli word
$X_0X_1Z_2Y_5Y_7$ would have flip indices of $(0, 1, 5, 7)$. A candidate generator $G_{i}$ would then be formed by taking one of the flip Pauli operators and changing the $X$ to $Y$ (or $Y$ to $X$) or one can think of applying a $S$ gate to the Pauli word. For the example Pauli word above, $G_i$ could be chosen as $Y_0X_1Y_5Y_7$. The energy lowering potential of the generator could then be found by utilizing the Rotosolve algorithm at Clifford points\cite{Ostaszewski2021, brown2025iterative}. This energy ordering produces better generators than using the gradient form\cite{brown2025iterative}. It was then noted in Ref \citenum{brown2025iterative} that one could deterministically obtain a Clifford transformation (composed of only CNOT gates) that converts the Hamiltonian to a form where a set of energy-lowering generators $G_{i}$ (acting only on single disjoint sets of qubits) was made possible.

To obtain this transformation, one sets up a binary matrix with the ``flip'' indices as rows and generators (ordered by energy lowering capability) as columns\cite{ryabinkin2020iterative}. For example, the highest energy contributing generators would have flip indices $(1,2,5,6),\, (0,2,5,7), (1,3,4,6), (0,2,4,6)$, $, \,(1,3,5,7)$ and others with smaller energy differences which are not linearly independent of the ones shown below. This corresponds to a matrix of
\begin{equation}\label{eq.g2diag}
    (\mathcal{G}|I)=\left[\begin{array}{cccccc|cccccccc}
        0&1&0&1&0&1&1&0&0&0&0&0&0&0\\ 
1&0&1&0&1&0&0&1&0&0&0&0&0&0\\ 
1&1&0&1&0&1&0&0&1&0&0&0&0&0\\ 
0&0&1&0&1&0&0&0&0&1&0&0&0&0\\ 
0&0&1&1&0&0&0&0&0&0&1&0&0&0\\ 
1&1&0&0&1&0&0&0&0&0&0&1&0&0\\ 
1&0&1&1&0&0&0&0&0&0&0&0&1&0\\ 
0&1&0&0&1&0&0&0&0&0&0&0&0&1
    \end{array}\right]
\end{equation}
one then concatenates an identity matrix to the right of the above array (denoted ($\mathcal{G}|I$) and performs Gaussian elimination over binary field to obtain $(G|T_{xx}^{T})$. This is shown in Eq. \ref{eq.diag}, where it can be seen that the flip indices for the first four operators are single $Y_{q}$ and the sixth generator is $Y_{4}$, and that there are no generators on qubits 5, 6 or 7. These are the expected three qubits that can be tapered\cite{bravyi2017taperingqubitssimulatefermionic}.

To define the transformation, we observe that there is a direct correspondence between the matrix $T_{xx}^{T}$ and the Clifford tableau representation\cite{Aaronson2004}. A Clifford tableau is a matrix representation used to efficiently track how Pauli operators transform under sequences of Clifford gates (such as CNOT, Hadamard, and phase gates). The tableau is typically organized into four quadrants, each encoding different aspects of these transformations. The top-left quadrant ($T_{xx}$) describes how X-type Pauli operators transform to X-type Pauli operators, while the top-right quadrant ($T_{xz}$) captures how Z-type information is transformed to X-type operators. Conversely, the bottom-left quadrant ($T_{zx}$) shows how X-type information influences the transformation of Z-type operators, and the bottom-right quadrant ($T_{zz}$) describes how Z-type operators transform under Z-type operations. CNOT gates only transform X-type to X-type and Z-type to Z-type with no cross operations so $T_{xz}=T_{zx}=0$.
$T_{xx}^{T}$ obtained from the Gaussian elimination defines the upper left quadrant of a standard Clifford tableau\cite{Aaronson2004} where the superscript $T$ denotes the transpose.

\begin{equation}
    (\mathcal{G}^{\prime}|T_{xx}^T)=\left[\begin{array}{cccccc|cccccccc}
        1&0&0&0&0&0&0&0&0&0&0&1&0&1\\ 
0&1&0&0&1&0&0&0&0&0&0&0&0&1\\ 
0&0&1&0&1&0&0&0&0&1&0&0&0&0\\ 
0&0&0&1&1&0&0&0&0&1&0&1&1&1\\ 
0&0&0&0&0&1&0&0&1&1&0&0&1&1\\ 
0&0&0&0&0&0&1&0&1&0&0&1&0&1\\ 
0&0&0&0&0&0&0&1&0&1&0&1&0&1\\ 
0&0&0&0&0&0&0&0&0&0&1&1&1&1\\ 
    \end{array}\right]\label{eq.diag}
\end{equation}

To determine the Clifford Tableau fully (or equivalently CNOT-only circuit) that transforms the Hamiltonian so that the highest contributing QCC generators are single $Y_{i}$, one simply performs Gaussian elimination over binary field of $(T_{xx}^{T}|I)->(A|T_{zz})$ to obtain $T_{zz}$. Then  $T_{xz}=T_{zx}=0$ which completely defines the Tableau that transforms the Pauli words in the Hamiltonian to other Pauli words. The number of terms in the Hamiltonian $\mathcal{O}(N^{4})$ with Clifford transformations scaling as $\mathcal{O}(N^{2})$\cite{nemirovsky2025reduced} for a total scaling of $\mathcal{O}(N^{6})$.

Before continuing, it is useful to see what type of transformation this performs on the corresponding unitary coupled-cluster\cite{Anand_2022} Fermionic operator.

\begin{equation}
    T_{ijkl} = \hat{a}^{\dagger}_i\hat{a}^{\dagger}_j\hat{a}_k\hat{a}_l-\hat{a}^{\dagger}_l\hat{a}^{\dagger}_k\hat{a}_j\hat{a}_i
\end{equation}

For example, the first generator from the example above is  $Y_{1}X_{2}X_{5}X_{6}$ which corresponds to one of the eight Pauli words that forms when the double excitation $a_{1}^{\dagger}a_{5}^{\dagger}a_2a_6-h.c.$ is mapped to qubits using the Jordan-Wigner mapping as shown in the second column of Table \ref{tab:placeholder}. After performing the transformation using the tableau derived above on the eight Pauli words, the corresponding qubit operator is
\begin{table}[]
    \centering
    \begin{tabular}{c|l|l}
    Majorana & JW Pauli & Transformed Pauli \\
    \hline
$1j (2, 4, 10, 13)$ & -1j Y1 X2 Y5 Y6 & +1j Y0 Z1 Z3 Z6 \\
$1j (2, 4, 11, 12)$ & 1j Y1 X2 X5 X6 & +1j Y0 Z2 Z6 \\
$-1j (2, 5, 10, 12$ & 1j Y1 Y2 Y5 X6 & -1j Y0 Z2 Z3 Z4 Z6 \\
$1j (2, 5, 11, 13)$ & 1j Y1 Y2 X5 Y6 & -1j Y0 Z1 Z4 Z6 \\
$-1j (3, 4, 10, 12)$ & -1j X1 X2 Y5 X6 & -1j Y0 Z1 \\
$1j (3, 4, 11, 13)$ & -1j X1 X2 X5 Y6 & -1j Y0 Z2 Z3 \\
$-1j (3, 5, 10, 13)$ & -1j X1 Y2 Y5 Y6 & +1j Y0 Z2 Z4 \\
$-1j (3, 5, 11, 12)$ & 1j X1 Y2 X5 X6 & +1j Y0 Z1 Z3 Z4 
    \end{tabular}
    \caption{The eight Majorana operators $\gamma_{2i}=a^{\dagger}_i+a_i, \gamma_{2i+1}=i(a_i^{\dagger}-a_i)$ of the double excitation operator $a_{1}^{\dagger}a_{5}^{\dagger}a_2a_6-h.c.$ converted to Pauli operators through the Jordan Wigner mapping and then transformed with Q-Cliff. You can see that after the transformation, the Paulis are $Z$ or $I$ operators on qubits 1 to 7 but always $Y$ on qubit $0$. }
    \label{tab:placeholder}
\end{table}
one which now only has the single $Y_0$ flip index on every mapped term as shown in the third column of Table \ref{tab:placeholder}.  Therefore, it is not unreasonable to say that we can approximate the double excitation ($a_{1}^{\dagger}a_{5}^{\dagger}a_2a_6-h.c.$) by performing the deterministic Clifford transformation and then utilizing a circuit $R_y(\theta_0)$ on qubit $0$. Likewise, the other single $R_Y$ generators correspond to different double (or single excitations) from the UCC ansatz. Finally, we observe that all terms in the Hamiltonian now have either $Z$ or $I$ on qubits $5$, $6$ and $7$. These are the three qubits that can be tapered (i.e. removed) where the initial state determines whether the term is multipled by $\pm1$ depending on the expectation value of the chosen initial state after transformation\cite{bravyi2017taperingqubitssimulatefermionic}. In that sense, one can think of the above procedure as a Hamiltonian inspired fermion-to-qubit mapping\cite{Parella_Dilm__2024}. For example, the generalized superfast encoding\cite{Setia_2019} would have many extra qubits that can be tapered due to the added stabilizers used for error mitigation/correction.  Although not examined explicitly here, it can matter which initial fermion-to-qubit mapping is used to generate the qubit Hamiltonian. Although the gradients should be the same regardless of the initial mapping, the order of the chosen generators can be different if terms have the same `exact' gradient but round off errors cause them to be different. 

\section{Classical Simulation}
\begin{figure}
     \centering
     \begin{subfigure}[b]{0.45\textwidth}
         \centering
         \includegraphics[width=\textwidth]{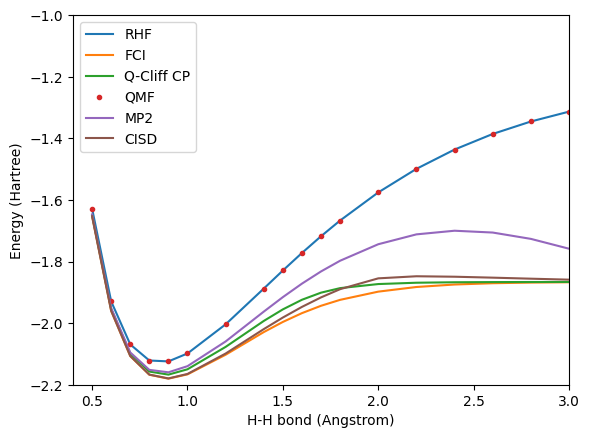}
         \caption{H$_4$}
         \label{fig:qmf_v_cp}
     \end{subfigure}
     \hfill
     \begin{subfigure}[b]{0.45\textwidth}
         \centering
         \includegraphics[width=\textwidth]{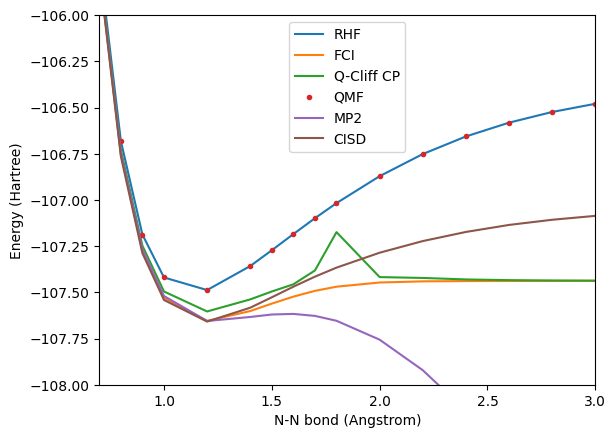}
         \caption{N$_2$}
         \label{fig:qmf_v_cp_n2}
     \end{subfigure}
        \caption{The importance of the Q-Cliff transformation for the H$_4$ and N$_2$ (with frozen core) systems in a STO-3G basis as a function of bond distance. As can be seen, the QMF wavefunction follows the RHF energy while the Q-Cliff CP wavefunction obtains energies that are better than MP2 for most of the potential energy curve. In the case of N$_2$, MP2 is better for short bond distances but the drawbacks or shortcomings of its non-variational nature becomes apparent quickly. CISD generally performs better for these molecules at short bond distances but its dissociation is worse due to the small basis set used.}
        \label{fig:three graphs}
\end{figure}

The transformation above converts the Hamiltonian to a form where single qubit $R_y(\theta_i)$ gates now have a gradient with respect to the energy which is not the case with most fermion-to-qubit mappings. We now examine what the simplest wavefunction approximation can achieve after Q-Cliff by variationally optimizing the $\theta_i$ angles. We note that although correlation (in the chemical sense) is introduced with the single $R_Y(\theta_i)$ ansatz, none of these single $R_y(\theta_i)$ gates introduces correlation between the qubits. One can think of this wavefunction approximation as a rank-1 canonical polyadic (CP) decomposition\cite{bro1997parafac} in qubit space. We can compare this wavefunction after the Clifford transformation to the standard QMF wavefunction produced in the original QCC paper\cite{ryabinkin2018qubitcoupledclustermethodsystematic,Ryabinkin_2018}. The QMF wavefunction is essentially the same (although $R_Z$ and $R_X$ are on each qubit) but starts from the standard Hartree-Fock mapped wavefunction. As can be seen in Figure \ref{fig:qmf_v_cp}, we immediately get more benefit by performing this simple (polynomial scaling) step. Whereas the QMF wavefunction obtains the RHF energy along the curve in these systems, the CP wavefunction obtains energies that are generally better than MP2.

\begin{figure}
    \centering
    \includegraphics[width=0.5\linewidth]{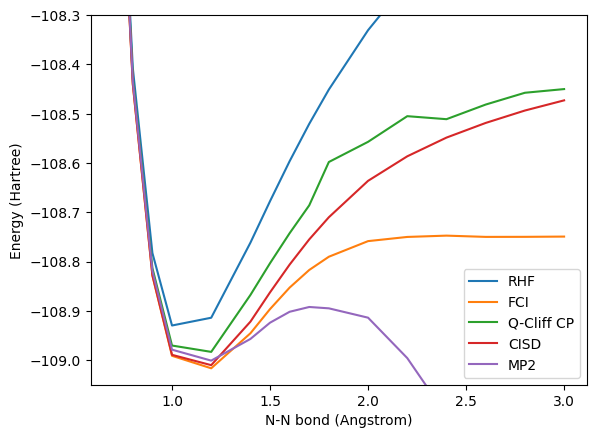}
    \caption{The CP wavefunction for a (8,8) active space for N$_2$ in a cc-pvqz basis. The CP wavefunction obtains an energy that is between MP2 and CISD accuracy.}
    \label{fig:placeholder}
\end{figure}

\subsection{DMRG entanglement}
\begin{figure}
     \centering
     \begin{subfigure}[b]{0.45\textwidth}
         \centering
         \includegraphics[width=\textwidth]{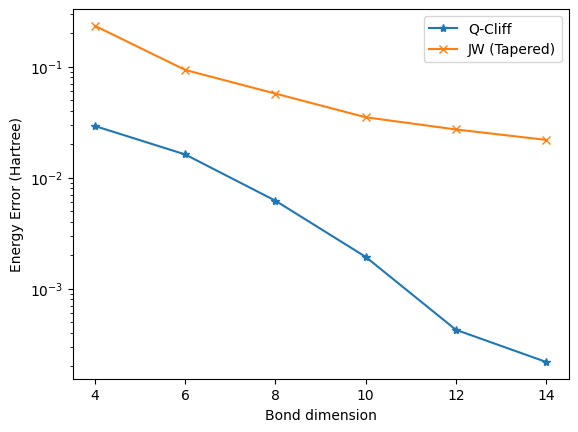}
         \caption{DMRG accuracy}
         \label{fig:dmrg}
     \end{subfigure}
     \hfill
     \begin{subfigure}[b]{0.45\textwidth}
         \centering
         \includegraphics[width=\textwidth]{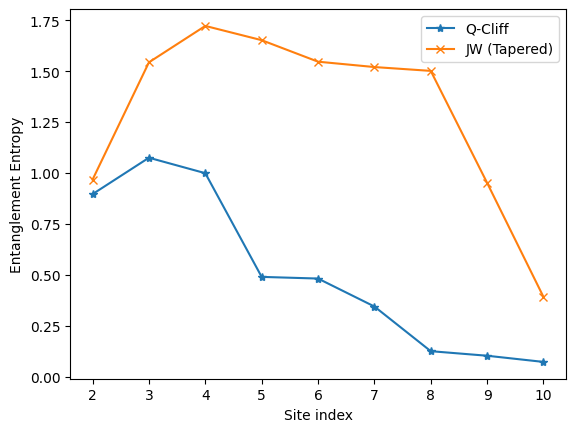}
         \caption{Entanglement Entropy (D=14)}
         \label{fig:ent}
     \end{subfigure}
        \caption{The increased efficiency of the DMRG calculations for the N$_2$ molecule at $2\AA$  using the cc-pVQZ basis in an (8,8) active space using the Q-Cliff transformed vs the tapered JW mapping. (a) The accuracy is far better after the Q-Cliff transformation as a function of bond distance. (b) The entanglement entropy between the DMRG qubit with a bond dimension of 14 sites is much smaller using the Q-Cliff derived mapping compared to the JW and tapered mapping.. }
        \label{fig:DRMG_N2}
\end{figure}

To further examine the properties of the fermion-to-qubit mapping using the Q-Cliff energy lowering capabilities, we perform a calculation on the N$_2$ molecule with DMRG and an increasing maximum bond dimension $D$. As can be seen in Figure \ref{fig:dmrg}, the accuracy of DMRG is far superior (approximately an order of magnitude better) using the Q-Cliff mapping compared to the JW mapping with the same number of qubits removed using standard transformations. To explain this, we examine the entanglement entropy between DMRG sites. If the entropy is lower, it should require a smaller bond dimension to be accurately represented. Figure \ref{fig:ent} shows that the entanglement entropy between qubits for the ground state has decreased by around half. It is also notable that the final three sites have minimal entanglement which provides support for the idea that the contributions there could be approximated using a perturbative approach\cite{Veis_2016}. To select the active space, one could utilize a method such as autocas\cite{stein2016automated}.

\section{Application to Quantum State Preparation}
In almost all cases, the CP wavefunction does not obtain chemical accuracy of $1$kcal/Mol so more excitations are required. However, we can utilize the uncorrelated non-Clifford wavefunction as a starting state to determine good generators $G_k$ and starting parameters $\theta_k$ by using previously established methods\cite{ryabinkin2019iterativequbitcoupledcluster}. By folding in all (or some) of the CP (i.e. $R_y(\theta_i)$) excitations, we also gain information about new generators that may not have been found given the RHF reference state. The process of folding in $R_y(\theta_i)$ rotations involves going term by term and applying the transformation for each Pauli $P$ in the Hamiltonian to generate one or two terms\cite{ryabinkin2019iterativequbitcoupledcluster}
\begin{equation}
    P+\sin[\theta_i]\left(-\frac{i}{2}\left[P,Y_i\right]\right)+\frac{1}{2}\left(1-\cos \theta_i\right)\left(Y_i P Y_i - P\right)
\end{equation}
depending on whether the operator on qubit $i$ commutes with $Y$.
This folding in process scales exponentially with the number of $R_y$ gates absorbed into the Hamiltonian but a subset could be folded in to expose more important generators or (in the case of chemistry) all singles, doubles, and triples could be efficiently screened for contributions. For large angles, one can also find the nearest Clifford angle $n\times \frac{\pi}{2}$ and use that to determine new generators.

In this manuscript, we fold in all the single-qubit excitations (generated by Q-Cliff) and determine the energy lowering capability of all possible generators from this new Hamiltonian. To obtain a short-depth circuit with good initial parameters, we decide ahead of time how many layers of a hardware efficient ansatz (HEA)\cite{ruhee2023challenges} that we wish to use. After ordering the generators from smallest-to-largest rotations, we can progressively build a Hardware efficient ansatz. 

The first step is to perform the same process as in Eq. \ref{eq.g2diag}. After performing the diagonalization, we obtain the tableau $T^{(1)}$ and remove the columns corresponding to those generator. This results in generators that are not diagonal as was initially the case. We again perform a Gaussian elimination with new matrix with fewer columns to obtain a new tableau ($T^{(2)}$). This process is repeated for the number of layers requested. To obtain a better initial state using the individually optimized parameters, we reverse the order of the operations such that large-angle operations are performed last in the circuit. This defines new $T^{(b)}$ by the indices of the possible parallel generators. So the circuit is compiled with 1) The CP gates followed by 2) the CNOT from tableau $T^{(n)}$ 3) a single layer of $R_y(\theta_i ^{(N)})$. Steps 2 and 3 are repeated with $n->n-1$ until $n=1$. We have now created a good representation of the wavefunction without any classical optimization. The process is deterministic and scales well with system size. 

\subsection{VQE Results}
\begin{figure}
     \centering
     \begin{subfigure}[b]{0.45\textwidth}
         \centering
         \includegraphics[width=\textwidth]{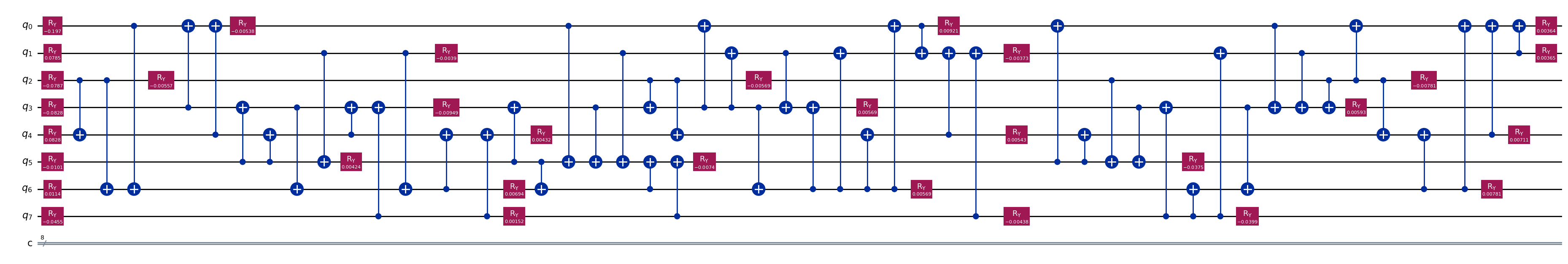}
         \caption{LiH Circuit}
         \label{fig:lih_circuit}
     \end{subfigure}
     \hfill
     \begin{subfigure}[b]{0.45\textwidth}
         \centering
         \includegraphics[width=\textwidth]{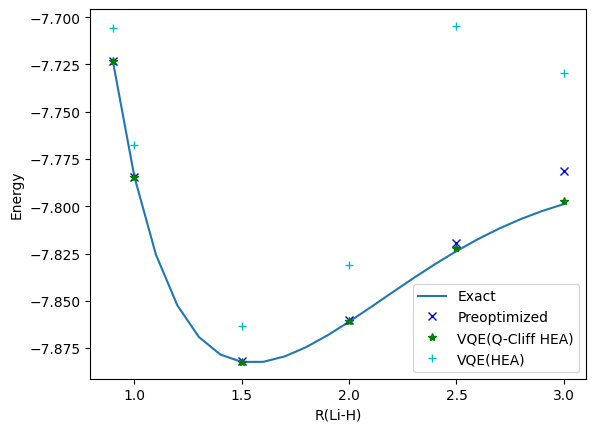}
         \caption{LiH PEC}
         \label{fig:lih_pec}
     \end{subfigure}
        \caption{The circuit and potential energy curve for a 3 layer Q-Cliff derived HEA for the LiH system. It can be seen that for short bond distances, excellent energies can be achieved using the Q-Cliff HEA preoptimization method. For larger bond distances, further optimization is required using VQE but a good representation of the state is obtained. The standard hardware efficient ansatz fails to obtain good results at any point on the curve.}
        \label{fig:lih_data}
\end{figure}

We first show that the method obtains a good starting state across the potential energy curve for LiH with $3$ layers. As can be seen in Figure \ref{fig:lih_pec}, the pre-optimized parameters get close to the exact result all along the potential energy curve. The system deterministically converges to below chemical accuracy when optimized. This is unlike the standard HEA which fails to achieve convergence as is well-known\cite{ruhee2023challenges}.

To show that the method can obtain more optimal parameters than state-of-the-art ADAPT methods, we show the comparison to obtain chemical accuracy with CEO-ADAPT-VQE in Table \ref{tab:ceo-v-qcc} for Li, BeH$_2$ and H$_6$ systems. As can be seen, the Q-Cliff derived HEA uses fewer CNOT gates and lower measurements but has higher gate depth at this point. Depending on the properties of the quantum computer, fewer CNOT gates may be more important than shorter depth. The measurement cost is lower because 1) we have a fixed ansatz that does not require multiple optimizations and 2) the starting guess for the parameters is reasonably close to the correct state. The increase in depth (compared to CEO-ADAPT-VQE) could probably be reduced by optimizing the CNOT layers\cite{PhysRevResearch.5.013065}. It should also be noted that the optimization was performed using SLSQP in Tangelo\cite{senicourt2022tangeloopensourcepythonpackage} without any fine tuning. It is certainly conceivable that a different optimizer could perform better and further reduce the number of measurements.

\begin{table}[]
    \centering
    \begin{tabular}{|c|c|c|c|c|}
        \hline
        Method & Metric & LiH at 3\AA & BeH$_2$ at 2\AA & H$_6$ at 1.5\AA \\
        \hline
        & CNOT Count & 107&218 & 812 \\
        CEO-ADAPT-VQE & Depth & 30& 95 & 282\\
        & Mesurement Count &560 & 2197 & 10857 \\
        \hline
        & CNOT Count & 51 (47\%) & 162 (74 \%) & N/A \\
        Q-Cliff HEA & Depth &37 (123\%) & 107 (112\%) & N/A\\
        & Mesurement Count & 463 (82\%) & 882 (40\%) & N/A\\
        \hline
    \end{tabular}
    \caption{A comparison of the Q-Cliff HEA ansatz vs CEO-ADAPT-VQE to obtain chemical accuracy for LiH, BeH$_2$ and H$_6$. The Q-Cliff HEA is generally better for moderately correlated systems, as it fails to achieve chemical accuracy for H$_6$. The percentage in brackets signifies the relative difference of Q-Cliff HEA to CEO-ADAPT-VQE.}
    \label{tab:ceo-v-qcc}
\end{table}

We also tested the Clifford transformation for the Benzene geometry given in Ref. \citenum{Parella_Dilm__2024}. For zero entangling layers, the Clifford transformation used here obtained an error of 31.13 kcal/mol and 14.40 kcal/mol after one entangling layer. The fermion-to-qubit mapping used in Ref \citenum{Parella_Dilm__2024} did not achieve 31.13 kcal/mol until around 6 or 8 entangling errors and never achieved 14.40 kcal/mol accuracy. Of course, our entangling layers are more complicated than the simple HEA ansatz used in Ref. \citenum{Parella_Dilm__2024} but for parameter efficient approximate state preparation, the Q-Cliff transformation seems to be a very good choice.

\section{Conclusions}
In this manuscript, we have evaluated the Q-Cliff transformation. We first show that it results in a cheap classical algorithm composed of a single-layer of $R_y(\theta_i)$ rotations that appears to have accuracy between MP2 and CISD. We also show that this method produces a fermion-to-qubit mapping that has less entanglement between the qubits which makes it far-more amenable to DMRG calculations. 

We expand on the techniques outlined in Ref \citenum{brown2025iterative} and show that we can generate short-depth circuits that prepare reasonable approximations to the ground state of various molecules. The resulting circuits are deterministically generated and provide a reasonable starting point for further quantum algorithms such as VQE and QPE. Using VQE, we show that the number of energy evaluations required to obtain chemical accuracy is competitive with state-of-the-art VQE methods. 

The first direction to take is using the fact that most of the entanglement between qubits is generally shifted to a smaller number of qubits. One could then obtain accurate results in this smaller space using DMRG or a quantum algorithm and then utilizing the RDMs to generate perturbative corrections outside of this space. A second direction is to combine the Clifford transformations derived here with DMRG similar to Ref \citenum{Qian_2024, Mello} but deterministically generating these Clifford transformations instead of sweeping over predefined patterns.

Given the above, the Hamiltonian derived Clifford transformation should be readily utilized for efficiently solving the ground states of quantum systems using both classical and quantum computers.

\bibliography{references}
\end{document}